\documentclass{article}

\usepackage{arxiv}
\usepackage[utf8]{inputenc}
 \usepackage{textcomp}
\usepackage[utf8]{inputenc} 
\usepackage[T1]{fontenc}    
\usepackage{hyperref}       
\usepackage{url}            
\usepackage{booktabs}       
\usepackage{amsfonts}       
\usepackage{nicefrac}       
\usepackage{microtype}      
\usepackage{lipsum}
\usepackage{graphicx}
\graphicspath{ {./images/} }

\title{Assessment of the January 2025 Los Angeles County Wildfires: A Multi-Modal Analysis of Impact, Response, and Population Exposure}

\author{
Seyd Teymoor Seydi \\
Department of Civil Engineering \\
Boise State University \\
Boise, ID 83706 USA \\
\texttt{seydseydi@boisestate.edu} \\
}

\begin{document}
\maketitle
\begin{abstract}

This study presents a comprehensive analysis of four significant California wildfires—Palisades, Eaton, Kenneth, and Hurst—examining their impacts through multiple dimensions, including land cover change, jurisdictional management, structural damage, and demographic vulnerability. Utilizing the Chebyshev-Kolmogorov-Arnold network (Cheby-KAN) model, applied to Sentinel-2 imagery, the extent of burned areas was mapped, ranging from 315.36 to 10,960.98 hectares. Our analysis revealed that shrubland ecosystems were consistently the most affected, comprising 57.4-75.8\% of burned areas across all events. The jurisdictional assessment demonstrated varying management complexities, ranging from singular authority (98.7\% in the Palisades Fire) to distributed management across multiple agencies. A subsequent structural impact analysis revealed significant disparities between urban interface fires (Eaton: 9,869 structures; Palisades: 8,436 structures) and rural events (Kenneth: 24 structures; Hurst: 17 structures). The demographic analysis revealed consistent gender distributions, with 50.9\% of the population identified as female and 49.1\% as male, across all events. The analysis also identified that working-age populations constituted the majority of the affected populations, ranging from 53.7\% to 54.1\%, and exhibited notable temporal shifts in post-fire periods. The study identified strong correlations between urban interface proximity, structural damage, and population exposure. The Palisades and Eaton fires affected over 20,000 people each, compared to fewer than 500 in rural events. These findings offer pivotal insights for the formulation of targeted wildfire management strategies, particularly in wildland-urban interface zones, and underscore the necessity for age- and gender-conscious approaches in emergency response planning.

\end{abstract}


\section{Introduction}

The accelerating impact of wildfires on human communities and infrastructure represents one of the most pressing challenges at the intersection of climate change and societal resilience. Recent years have witnessed an unprecedented surge in wildfire activity across the western United States, with California experiencing a dramatic fivefold increase in annual burned area during 1972-2018 \cite{Williams2019}. This escalation, punctuated by record-breaking fire seasons in 2017 and 2018, has raised urgent questions about the vulnerability of human populations and critical infrastructure in fire-prone regions \cite{Modaresi2023}.

The growing wildfire crisis emerges from a complex interplay of factors. Climate change has fundamentally altered fire regimes through warming-driven increases in atmospheric aridity, with warm-season temperatures rising approximately 1.4°C since the early 1970s \cite{Bowman2017}. This warming has occurred against a backdrop of expanding human development into fire-prone areas, with the wildland-urban interface (WUI) growing significantly over recent decades \cite{Radeloff2018}. The convergence of these trends has created unprecedented challenges for fire management and community protection.

Recent research has revealed concerning patterns in human and infrastructure exposure to wildfires. Cumulative primary human exposure—defined as the population residing within the perimeters of large wildfires—reached 594,850 people during 2000-2019 across the contiguous United States, with 82\% concentrated in the western regions \cite{Modaresi2023}. This exposure shows marked regional variation, with California accounting for a disproportionate share of both population exposure and infrastructure vulnerability. The state has experienced widespread exposure of critical infrastructure, including 412,155 km of roads and 14,835 km of transmission powerlines \cite{Modaresi2023}.

The relationship between climate change and wildfire activity demonstrates notable seasonal and regional heterogeneity. Summer forest fires show the strongest correlation with warming-induced aridity, while fall wildfires often result from the interaction between strong offshore wind events and dry fuels \cite{Williams2019}. This seasonal variation combines with regional differences in vegetation, topography, and human development patterns to create distinct spatial patterns of fire risk and vulnerability.

Despite a growing understanding of these patterns, significant knowledge gaps remain in our understanding of how wildfires differentially affect different communities and infrastructure across diverse landscapes. While previous studies have examined singular aspects of wildfire exposure, a comprehensive analysis that integrates population demographics, infrastructure vulnerability, and land cover change across multiple fire events is lacking. In addition, the use of advanced remote sensing techniques for accurate burn area mapping in complex terrain remains underexplored, particularly in the context of varying jurisdictional complexities.

To address these gaps, this study presents a multi-dimensional analysis of four significant California wildfires—Palisades, Eaton, Kenneth, and Hurst—examining their impacts through multiple analytical lenses. Specifically, this research aims to:
\begin{enumerate}
    \item Quantify and compare burn patterns and land cover changes across urban and rural fire events using advanced remote sensing techniques
    \item Analyze jurisdictional complexities in fire management and their implications for emergency response
    \item Assess structural vulnerability patterns at the wildland-urban interface
    \item Evaluate demographic characteristics and temporal shifts in affected populations
\end{enumerate}

The primary contributions and novelty of this research are fourfold:
\begin{enumerate}
    \item Methodological Innovation: Introduction of the Chebyshev-Kolmogorov-Arnold network (Cheby-KAN) \cite{ss2024chebyshev,Seydi2024} model for wildfire mapping, providing enhanced accuracy in burn area assessment through Sentinel-2 imagery analysis. This novel approach enables precise quantification of fire extents ranging from 315.36 to 10,960.98 hectares.
    
    \item Comparative Framework: Development of a comprehensive analytical framework that enables direct comparison between urban interface and rural fire events, revealing previously undocumented disparities in structural impact (ranging from 17 to 9,869 affected structures) and population exposure (from under 500 to over 20,000 people).
    
    \item Demographic Vulnerability Assessment: Novel integration of demographic analysis with fire impact assessment, revealing consistent patterns in gender distribution (50.9\% female, 49.1\% male) and working-age population predominance (53.7-54.1\%) across diverse fire events, contributing to more targeted emergency response planning.
    
    \item Management Complexity Analysis: First systematic assessment of jurisdictional complexity in California fire management, ranging from single-authority scenarios (98.7\% in Palisades) to multi-agency frameworks, providing crucial insights for emergency response coordination.
\end{enumerate}

\section{Study Area and Dataset}
\subsection{Study Area}

\begin{figure}[h!]
    \centering
    \includegraphics[width=\textwidth]{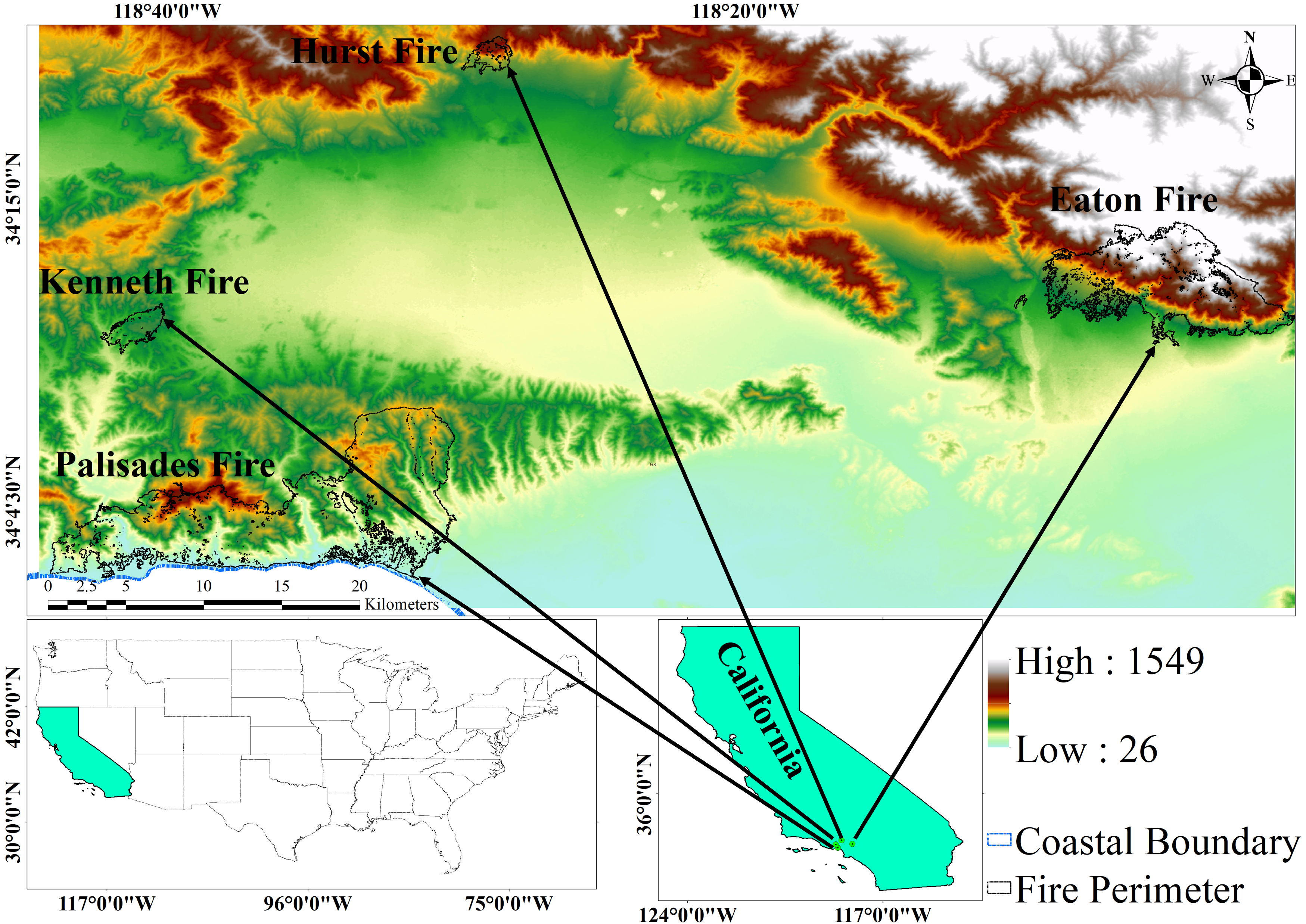}
    \caption{Geographic location of the four major wildfire events in Los Angeles County, California.}
    \label{fig:study_area}
\end{figure}

The present study focused on four significant wildfire events that occurred in Los Angeles County, California, during January 7-10, 2025: the Palisades, Eaton, Kenneth, and Hurst Fires (Figure \ref{fig:study_area}). The study area encompasses a topographically complex region, with elevations ranging from 26 to 1,549 meters above sea level. This area is characterized by diverse terrain, including coastal zones, urban interfaces, and mountainous landscapes.

The study area's intricate jurisdictional structure necessitated a coordinated response effort by multiple agencies. According to the Congressional Research Service report, the majority of land in Los Angeles County is classified as having Very High, High, or Moderate wildfire hazard potential. This classification, in conjunction with the region's escalating urbanization in wildland-urban interface zones, where nearly a third of residences are situated, engenders particularly challenging conditions for fire management and community protection.

\subsection{Datasets}

The study utilized multiple high-resolution geospatial datasets to conduct a comprehensive assessment of wildfire impacts. The primary demographic analysis was based on WorldPop Global Project data \cite{sorichetta2015americas,linard2012africa,gaughan2013asia}, which provides population distribution and age-sex structure information at 100m resolution. These datasets employ machine learning approaches to disaggregate census-based population counts, offering detailed demographic insights for 2020-2021. The population data was accessed through Google Earth Engine (\url{ee.ImageCollection("WorldPop/GP/100m/pop_age_sex"}) and \url{ee.ImageCollection("WorldPop/GP/100m/pop")}), facilitating efficient large-scale analysis.

Land cover analysis was performed using the USGS National Land Cover Database 2021 (NLCD), which provides national land cover classification at 30m resolution \cite{dewitz2023nlcd}. This dataset, available through Google Earth Engine (\url{ee.ImageCollection("USGS/NLCD_RELEASES/2021_REL/NLCD")}), was essential for understanding the distribution of different land cover types affected by the fires and assessing changes in urban-wildland interface zones.

Protected area analysis utilized the Protected Areas Database of the United States (PAD-US v2.0), accessed via Google Earth Engine (\url{ee.FeatureCollection("USGS/GAP/PAD-US/v20/fee")}). This database serves as America's official inventory of protected areas, providing crucial information about land management and conservation status \cite{usgs2018padus}. The dataset offers comprehensive coverage of areas dedicated to preserving biological diversity, natural resources, and cultural uses.

Building infrastructure assessment was conducted using California Building Footprints data, which provides high-resolution building infrastructure information derived from remote sensing imagery \cite{dao2020california}. This dataset was essential for quantifying structural impacts and analyzing patterns of fire damage in developed areas (\url{https://zenodo.org/records/3979706}).

Pre- and post-fire conditions were analyzed using Sentinel-2 multispectral imagery, providing 10m resolution data across multiple spectral bands (B2-B8, B11-B12). The imagery was carefully selected to minimize cloud cover (<10\%) and ensure optimal temporal proximity to the fire events, enabling detailed burned area mapping and change detection analysis (\url{ee.ImageCollection("COPERNICUS/S2_HARMONIZED")}).

\section{Methodology}

\subsection{Burned Area Detection Framework}
\begin{figure}[h!]
\centering
\includegraphics[width=\textwidth]{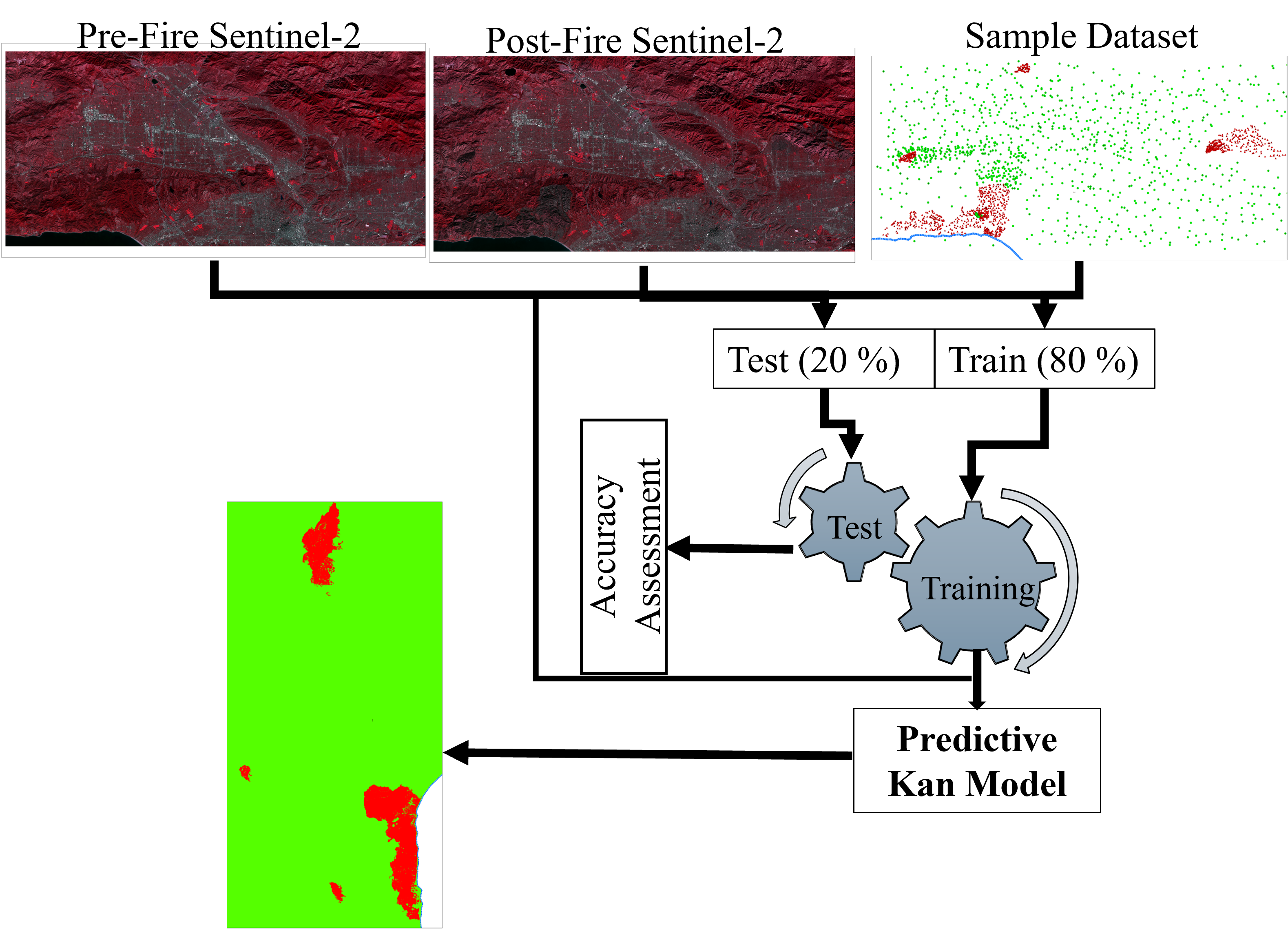}
\caption{Workflow for burned area detection using Sentinel-2 imagery and KAN model.}
\label{fig:methodology_workflow}
\end{figure}

Our methodology employs a sophisticated approach integrating Sentinel-2 satellite imagery with the Chebyshev-Kolmogorov-Arnold network (Cheby-KAN) \cite{ss2024chebyshev} model for precise burned area detection (Figure \ref{fig:methodology_workflow}). Pre- and post-fire Sentinel-2 imagery was acquired through the Google Earth Engine (GEE) platform, with rigorous selection criteria in place to ensure the integrity of the data \cite{Seydi2022}. These criteria included stringent cloud cover thresholds (<10\%) and optimal temporal proximity to fire events. The multispectral data stack incorporated ten spectral bands (B2-B8, B11-B12) at 10-meter spatial resolution, enabling detailed surface characteristic analysis. To ensure a robust training dataset, a stratified random sampling strategy was implemented, ensuring representative coverage across burned (red) and unburned (green) areas. The dataset was partitioned using an 80:20 ratio for training and testing, respectively, maintaining class proportions to prevent bias. The Cheby-KAN model architecture underwent optimization through iterative training procedures, incorporating dropout layers (rate = 0.3) to prevent overfitting and batch normalization to accelerate training convergence. The performance of the model was evaluated using a held-out test dataset, with accuracy metrics including overall accuracy, the Kappa coefficient, and class-specific F1 scores. The validated model was then applied to generate binary burned area masks, with post-processing including morphological operations to reduce noise and enhance spatial coherence.

\subsection{Impact Assessment Framework}
\begin{figure}[h!]
\centering
\includegraphics[width=\textwidth]{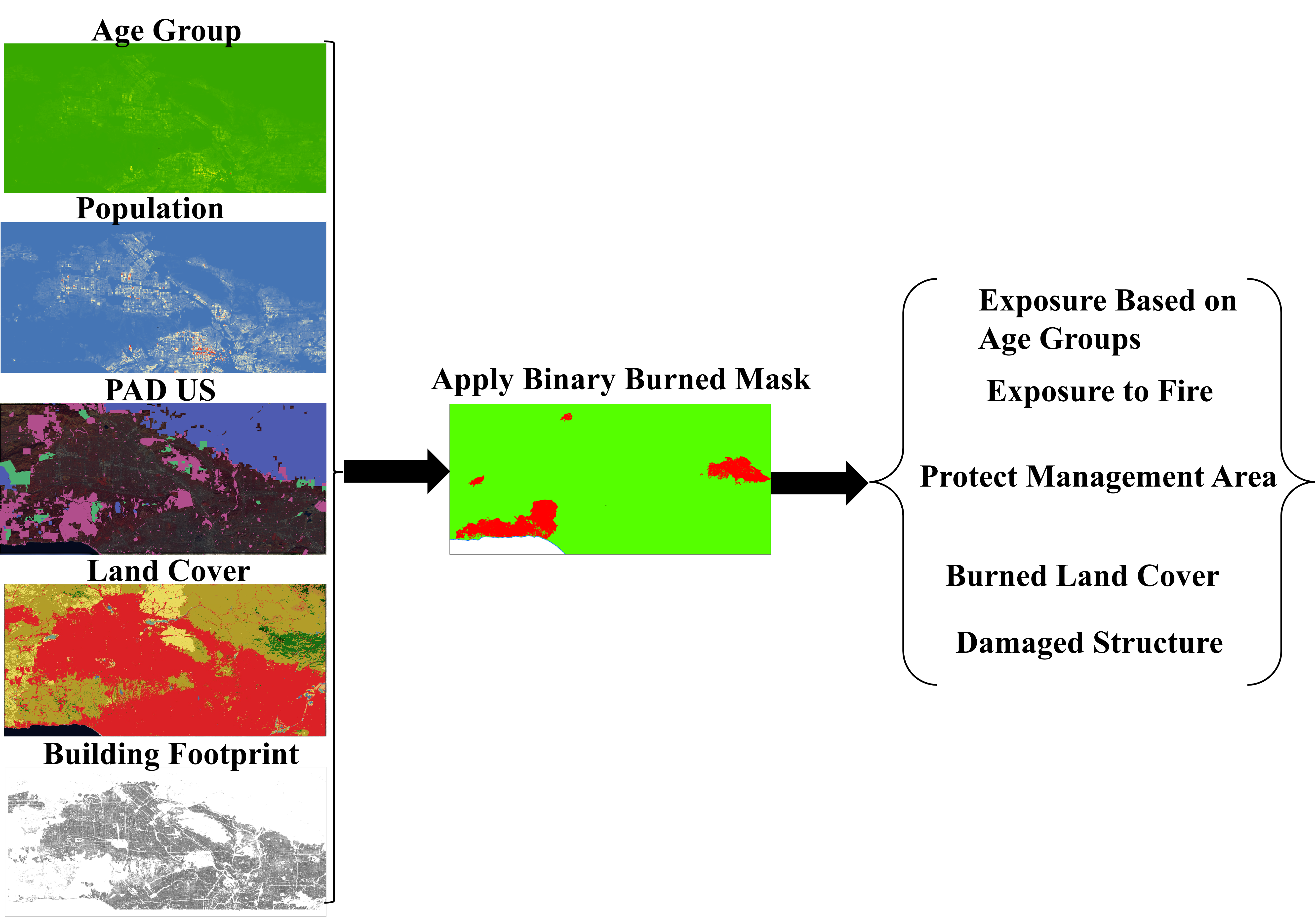}
\caption{Multi-criteria framework for fire impact assessment.}
\label{fig:impact_framework}
\end{figure}

The impact assessment framework (Figure \ref{fig:impact_framework}) integrates multiple geospatial datasets to provide a comprehensive analysis of fire impacts across social, ecological, and infrastructure dimensions. The framework synthesizes five key input datasets: WorldPop age/sex distribution data (100m resolution) for demographic analysis, WorldPop population count data validated against census benchmarks, Protected Areas Database of the United States (PAD-US Version 3.0) for management and protection status, high-resolution (30m) land use/land cover classification with focus on urban-wildland interface zones, and building footprint data derived from high-resolution imagery and LiDAR datasets. The binary burned mask was systematically applied to these input datasets using Geographic Information System (GIS) overlay analysis. Spatial statistics were computed using moving window operations to capture local variations in impact patterns, while dasymetric mapping techniques were employed to improve population distribution estimates. The analytical framework that was employed is capable of generating comprehensive outputs, including but not limited to the following: age-stratified exposure analysis, population exposure quantification, protected area impact assessment, land cover analysis, and infrastructure impact assessment.

\section{Results}

\subsection{Burned Area Mapping}
\begin{figure}[h!]
    \centering
    \includegraphics[width=\textwidth]{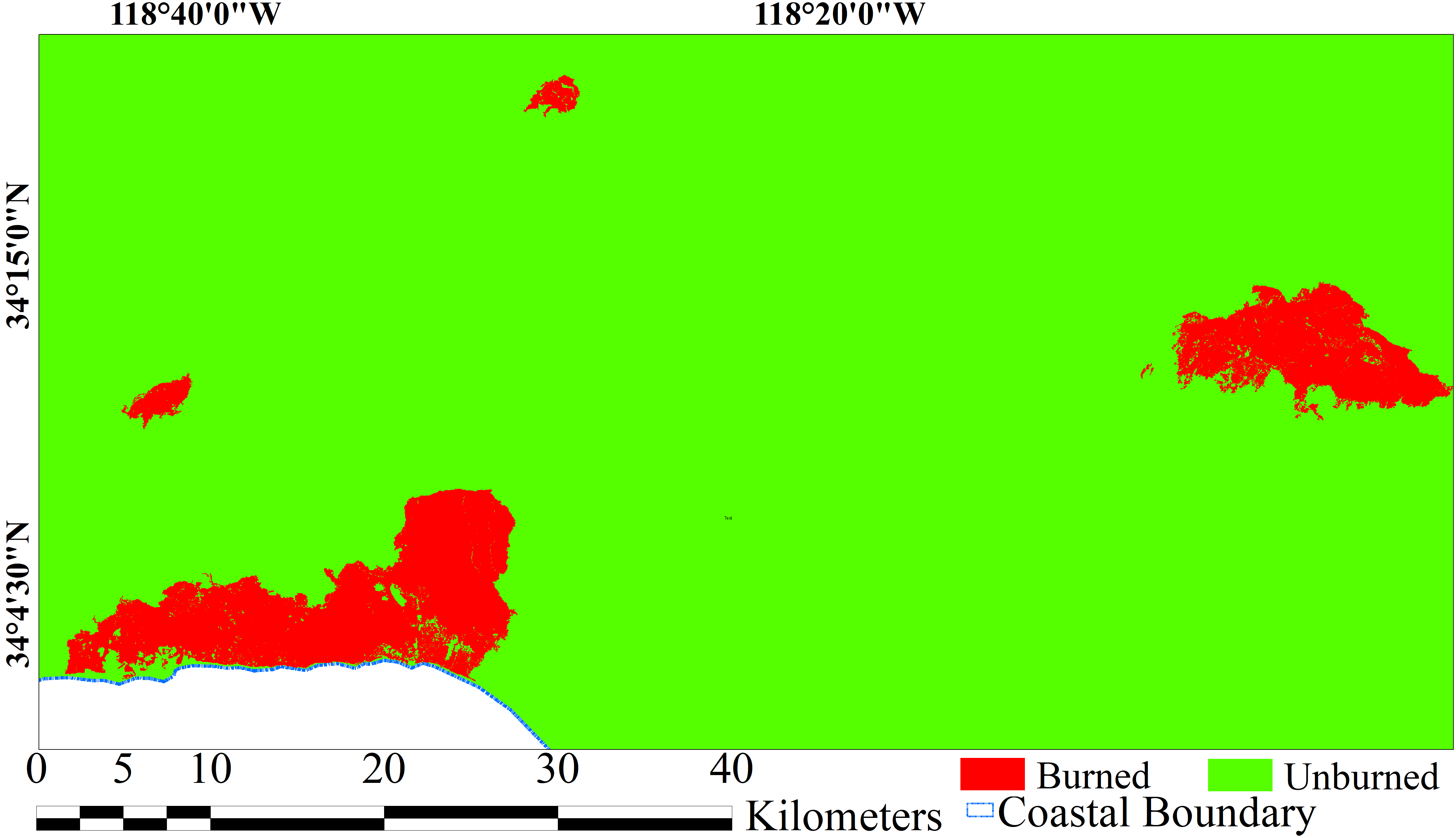} 
    \caption{Burned area mapping results obtained using the Kolmogorov-Arnold Network (KAN) model, applied to Sentinel-2 imagery. The map highlights burned areas (red) and unburned areas (green), with coastal boundaries marked in blue dashed lines. The masking of noisy label pixels in unburned regions enhanced the accuracy and reliability of the results.}
    \label{result_bam}
\end{figure}

Figure \ref{result_bam} illustrates the burned area mapping results obtained using the Kolmogorov-Arnold network (KAN) model, specifically the Chebyshev-KAN (Cheby-KAN) architecture, applied to Sentinel-2 imagery. The map effectively distinguishes between burned (red) and unburned (green) regions, with coastal boundaries (blue dashed lines) providing spatial context. To improve the accuracy and reliability of the results, noisy label pixels in unburned areas were removed through a masking process. This approach significantly reduced false positives, resulting in better delineation of burned areas and more accurate analysis. The model successfully mapped four distinct fire events: the Hurst Fire (315.36 ha), Eaton Fire (5,325.77 ha), Kenneth Fire (440.74 ha), and Palisades Fire (10,960.98 ha), with a total combined burned area of 17,042.85 hectares.

The model demonstrated high performance on the test dataset, achieving an overall accuracy of 98.68\%, a kappa coefficient of 0.9714, and an F1 score of 0.9817. These metrics indicate the robustness of the Cheby-KAN model in identifying fire-affected regions with minimal errors. The high kappa value underscores the strong agreement between predicted and true classifications, while the F1 score reflects a balanced performance in handling both precision and recall. The removal of noisy labels further contributed to reducing misclassifications, ensuring a clear depiction of burned areas suitable for decision-making and environmental assessment.

\begin{table}[h!]
\centering
\caption{Performance metrics of the Cheby-KAN model on the test dataset.}
\label{tab:performance_metrics}
\begin{tabular}{|c|c|}
\hline
\textbf{Metric}          & \textbf{Value} \\ \hline
Overall Accuracy         & 0.9868         \\ \hline
Kappa Coefficient        & 0.9714         \\ \hline
F1 Score                 & 0.9817         \\ \hline
\end{tabular}
\end{table}

\subsection{Burned Land Cover Land Use Estimation}

A comprehensive analysis of land cover and land use (LCLU) distribution across four major fire events was conducted, unveiling distinct patterns of fire propagation and ecosystem vulnerability. The spatial distribution of burned areas exhibited a marked susceptibility of shrubland ecosystems, with notable variations in the impact on urban-wildland interfaces and forest systems.
The most salient pattern pertained to the predominance of shrubland ecosystems as primary fuel sources, with proportions ranging from 57.4\% to 75.8\% across all events. Specifically, Tropical Shrubland exhibited the highest vulnerability in three of the four fires (Eaton, Kenneth, and Palisades), while Temperate Shrubland dominated in the Hurst Fire (63.4\%). This consistent pattern suggests that shrubland ecosystems, regardless of their climatic classification, serve as principal drivers of fire propagation in these landscapes.

A notable observation is the presence of distinct secondary burn patterns, indicative of local ecosystem heterogeneity, exhibited by the fires. The Hurst Fire, for instance, exhibited a substantial impact on Mixed Forest (26.6\%), indicating a complex fire behavior at the shrubland-forest ecotone. Conversely, the Eaton Fire exhibited notable involvement from the urban interface (19.0\%) along with substantial impact on the Tropical Broadleaf Evergreen forest (18.7\%), suggesting a multifaceted urban-wildland fire dynamic.
The Kenneth fire offered a distinctive illustration of ecosystem interaction, in which Tropical Grassland contributed a significant portion (40.1\%) to the total burned area, in conjunction with Tropical Shrubland (57.4\%). This phenomenon suggests the presence of a grass-fire cycle dynamic, where the coexistence of both fuel types may have exerted an influence on the behavior and spread patterns of the fire.

The Palisades Fire, which was the most extensive in terms of area, exhibited the highest proportion of involvement by Tropical Shrubland (75.8\%) among all events, with significant urban interface burning (15.6\%). This observation suggests that landscape homogeneity may play a pivotal role in facilitating the spread of fires, particularly in the presence of urban proximity.

The consistent presence of an "Other" land cover category, ranging from 0. 3\% to 4. 9\%, was observed in all fires, indicating the complexity of landscape matrices in fire-prone regions. This heterogeneity in landscape composition carries significant ramifications for the formulation of fire management strategies and the development of post-fire recovery plans.

These findings underscore the pivotal role of vegetation type and landscape configuration in dictating fire behavior and extent. The consistent vulnerability of shrubland ecosystems, coupled with varying degrees of involvement from urban interfaces, underscores the necessity for ecosystem-specific fire management approaches and careful consideration of urban-wildland boundaries in fire risk assessments.

\begin{figure}[h!]
    \centering
    \includegraphics[width=\textwidth]{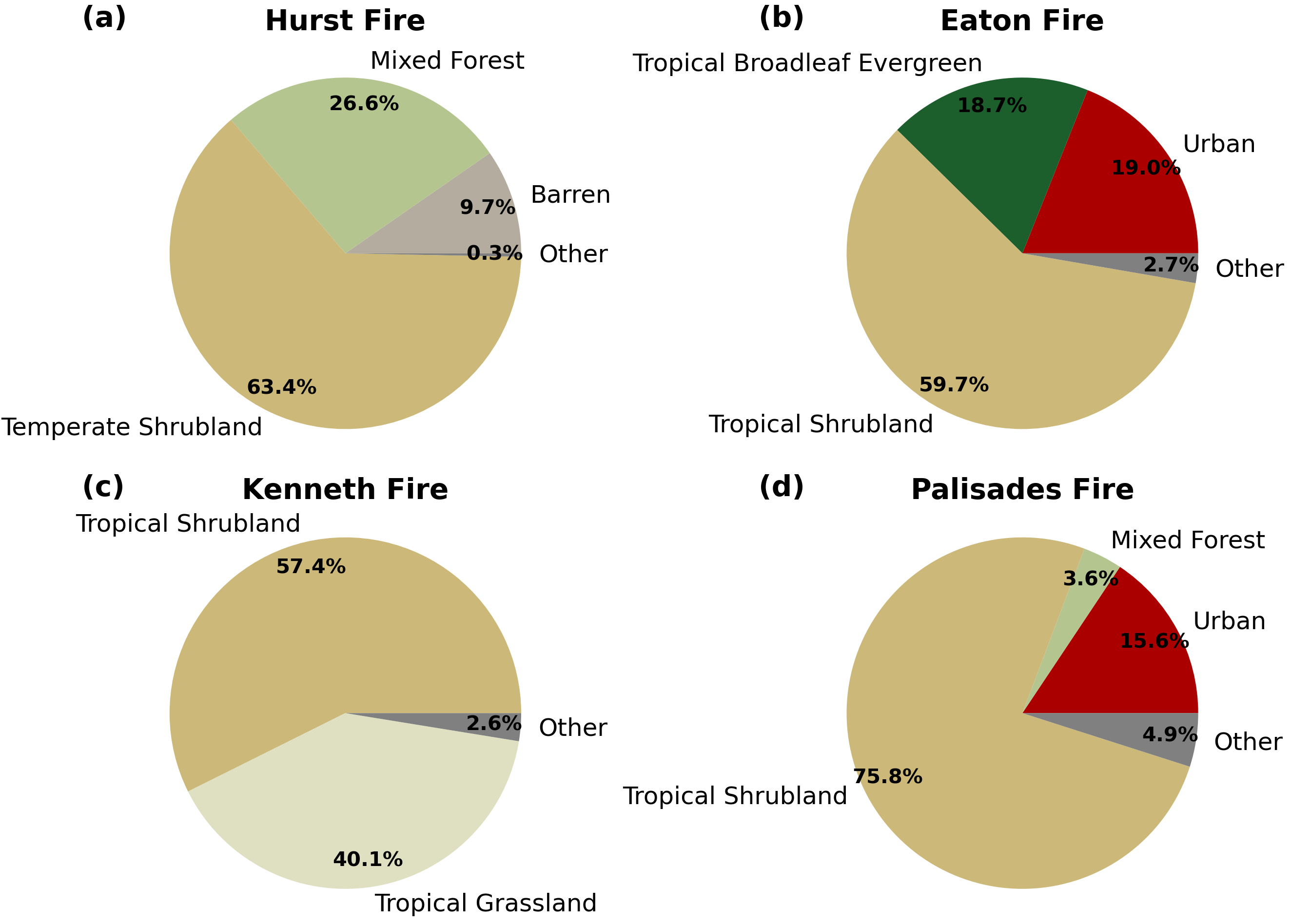} 
    \caption{Burned land cover and land use distribution for four fire events: (a) Hurst Fire, (b) Eaton Fire, (c) Kenneth Fire, and (d) Palisades Fire. }
    \label{fig:fire_distribution}
\end{figure}

\subsection{Building Damage Assessment and Spatial Distribution}
\begin{figure}[h!]
    \centering
    \includegraphics[width=0.8\textwidth]{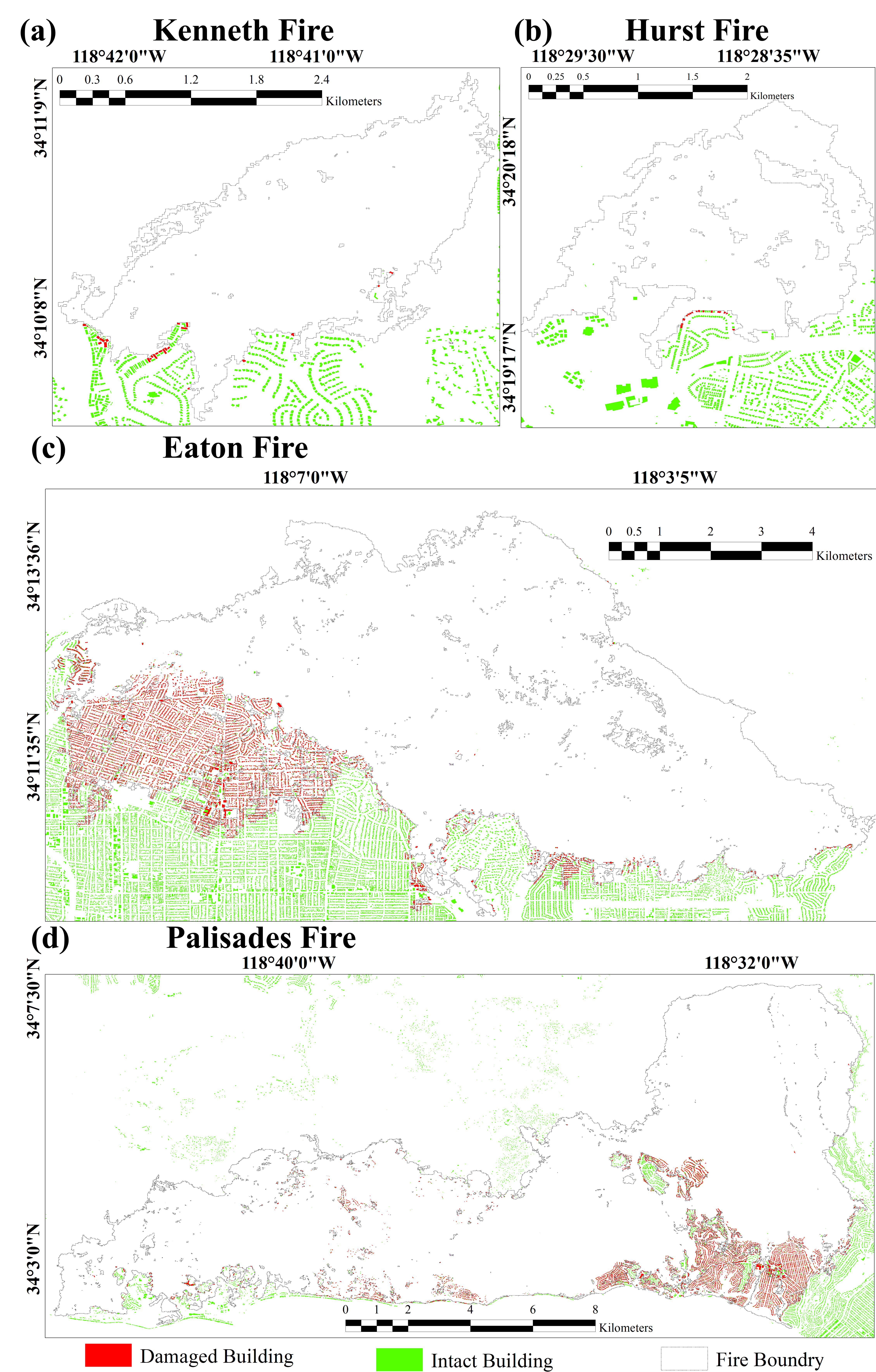} 
    \caption{Number and spatial distribution of damaged buildings for four fire events: (a) Kenneth Fire, (b) Hurst Fire, (c) Eaton Fire, and (d) Palisades Fire.}
    \label{fig:damage_distribution}
\end{figure}

The spatial analysis of building damage patterns across the four wildfire events revealed significant variations in both the magnitude and distribution of structural impacts (Figure \ref{fig:damage_distribution}). Our comprehensive assessment utilizing high-resolution Sentinel-2 imagery and the Chebyshev-KAN model identified distinct patterns of building damage that correlate with urban-wildland interface configurations and local topography.

The Eaton Fire emerged as the most destructive event, resulting in 9,869 damaged or destroyed structures. The spatial distribution map (Figure \ref{fig:damage_distribution}c) shows a concentrated pattern of structural damage (red) predominantly in the western portion of the fire perimeter, with intact buildings (green) forming a clear boundary around the impact zone. This pattern suggests an intense urban interface fire that rapidly spread through developed areas, possibly driven by local wind patterns and urban fuel connectivity.

The Palisades Fire caused the second-highest structural impact with 8,436 damaged buildings. The damage distribution (Figure \ref{fig:damage_distribution}d) exhibits a distinctive east-west elongated pattern along the coastline, with clusters of damaged structures interspersed among intact buildings. This spatial arrangement indicates a complex fire behavior influenced by coastal topography and the urban-wildland interface configuration.

In contrast, the Kenneth and Hurst Fires showed significantly lower structural impacts, with 24 and 17 damaged buildings respectively. The Kenneth Fire's damage pattern (Figure \ref{fig:damage_distribution}a) shows sparse and isolated structural impacts at the urban periphery, suggesting effective containment strategies or natural fire breaks that limited spread into developed areas. Similarly, the Hurst Fire (Figure \ref{fig:damage_distribution}b) displays minimal structural damage, with affected buildings primarily located at the southern edge of the fire perimeter.

The spatial analysis reveals a notable correlation between building density and damage extent. Areas with higher building density, particularly in the Eaton and Palisades Fire zones, experienced more substantial structural losses, highlighting the critical role of urban configuration in fire vulnerability. The clear delineation between damaged and intact structures, particularly visible in the Eaton Fire map, provides valuable insights for urban planning and fire mitigation strategies in wildland-urban interface zones.

These findings underscore the importance of understanding spatial patterns of fire-induced structural damage for improving wildfire risk assessment and urban planning in fire-prone regions. The stark contrast in damage patterns between high-density urban interface fires (Eaton, Palisades) and more rural events (Kenneth, Hurst) emphasizes the need for location-specific fire management strategies that account for local building patterns and urban-wildland interface characteristics.

\begin{table}
 \caption{Number of structures damaged or destroyed by wildfires in California for each study area.}
  \centering
  \begin{tabular}{|l|c|}
    \hline
    \textbf{Fire Name} & \textbf{Number of Structures Damaged or Destroyed} \\
    \hline
    Eaton          & 9869 \\
    Palisades      & 8436 \\
    Hurst          & 17   \\
    Kenneth        & 24   \\
    \hline
  \end{tabular}
  \label{tab:fire_damage}
\end{table}

\subsection{Protected Agency Jurisdiction Analysis}

\begin{figure}[h!]
    \centering
    \includegraphics[width=\textwidth]{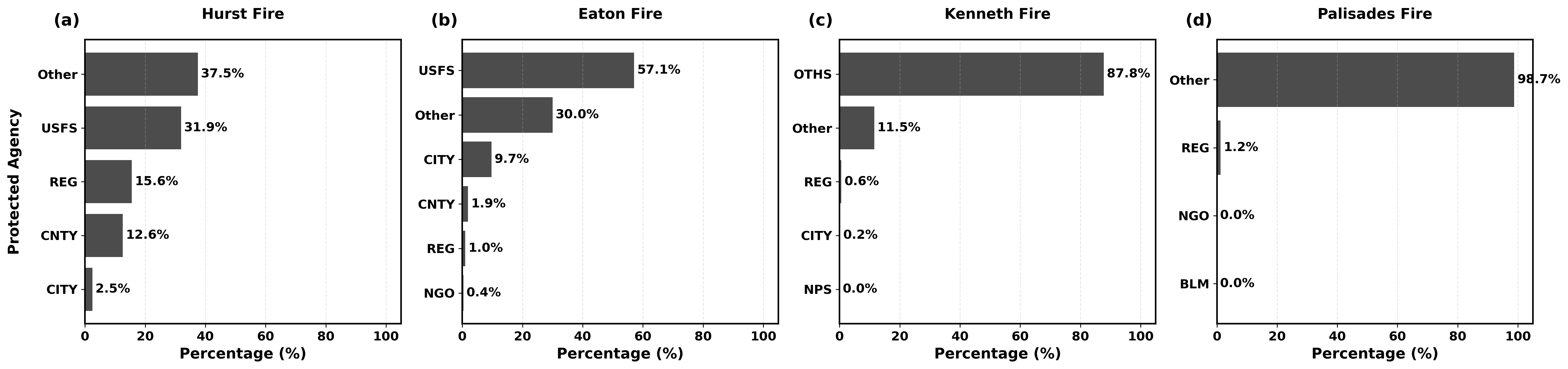}
    \caption{Distribution of protected agency jurisdictions across four California wildfire events.}
    \label{fig:protected_area_distribution_bar}
\end{figure}

An analysis of protected agency jurisdictions across the four wildfire events revealed distinct patterns of land management responsibility and jurisdictional complexity (Figure \ref{fig:protected_area_distribution_bar}. Each fire exhibited a unique distribution of agency oversight, reflecting the diverse landscape of fire management authorities in California.
The Palisades Fire exhibited the most concentrated jurisdictional pattern, with 98.7\% of the affected area falling under "Other" jurisdiction, while Regional Agencies (REG) maintained minimal oversight at 1.2\%. This near-complete dominance of a single jurisdictional category suggests streamlined emergency response coordination, though it may also indicate potential challenges in cross-agency resource mobilization.
A similar yet distinct jurisdictional pattern was observed in the Kenneth Fire, with Other Health Services (OTHS) overseeing 87.8\% of the affected area. The remaining jurisdictions were distributed among "Other" (11.5\%), Regional agencies (0.6\%), and City authorities (0.2\%), indicating a landscape predominantly managed by health services with minimal multi-agency involvement.
In contrast, the Eaton Fire exhibited more complex jurisdictional divisions, with the United States Forest Service (USFS) managing the majority (57.1\%) of the affected area. The remaining 30.0\% of the affected area fell under the jurisdiction of "Other" authorities, while City (9.7\%), County (CNTY, 1.9\%), Regional (1.0\%), and Non-Governmental Organizations (NGO, 0.4\%) each accounted for smaller portions. This intricate jurisdictional landscape underscores the necessity for enhanced inter-agency coordination during fire response operations.
The Hurst Fire exhibited the most balanced jurisdictional distribution among the four events. "Other" jurisdictions managed 37.5\% of the area, followed closely by USFS at 31.9\%. Regional agencies exercised control over 15.6\% of the area, followed by county agencies at 12.6\%, and city authorities with 2.5\%. This relatively balanced distribution indicates a complex management landscape requiring significant inter-agency cooperation during fire response.
These jurisdictional patterns carry significant ramifications for the formulation of fire management strategies, the coordination of emergency responses, and the undertaking of post-fire recovery efforts. The varying degrees of jurisdictional complexity across these fires underscore the necessity for robust inter-agency communication protocols and coordinated response frameworks, particularly in areas with multiple managing authorities.

\subsection{Population Demographics and Exposure Analysis}

\begin{figure}[h!]
    \centering
    \includegraphics[width=\textwidth]{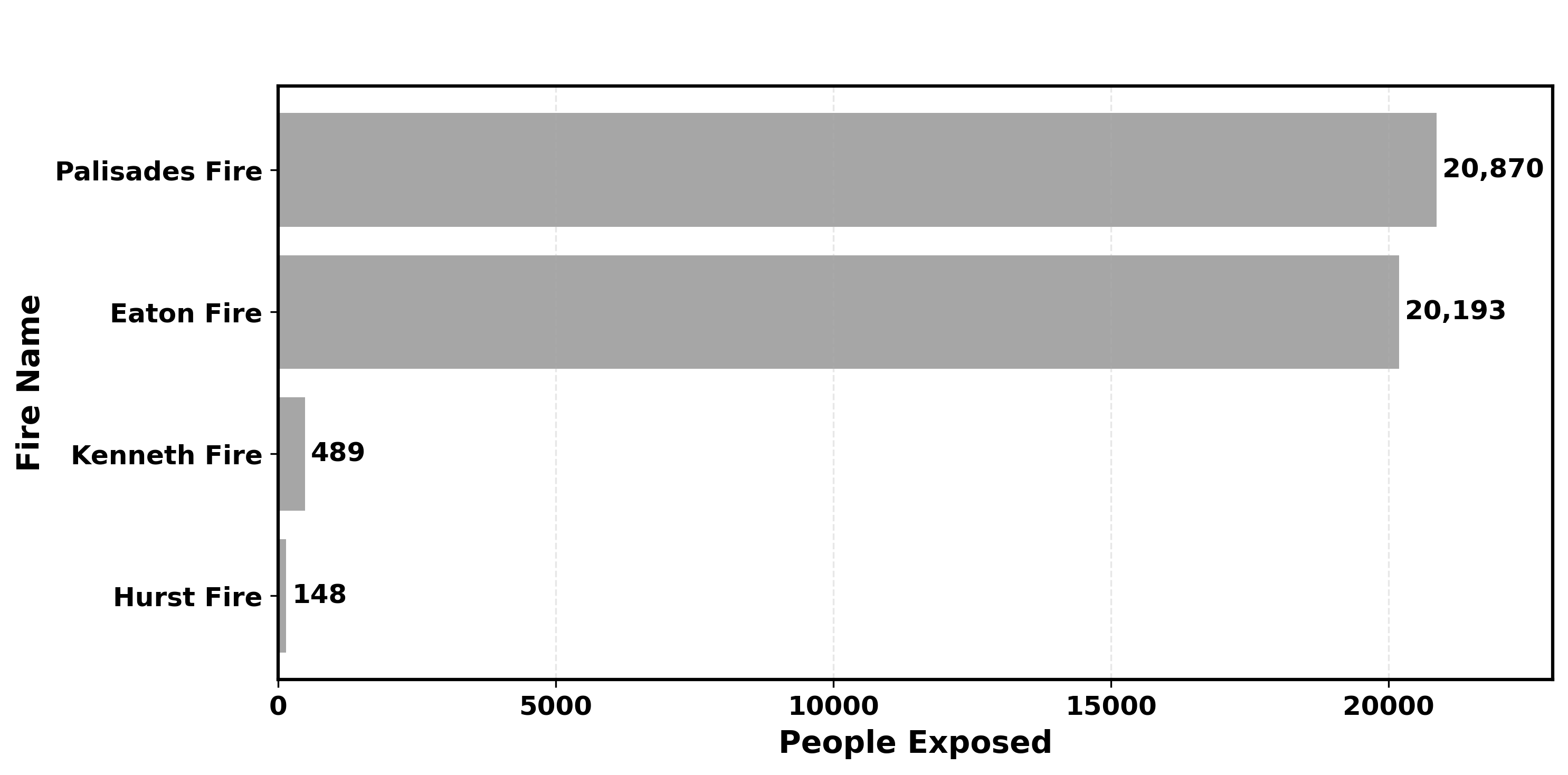}
    \caption{Population exposure analysis for four wildfire events in California, showing the total number of people exposed to each fire event. }
    \label{fig:pop_exposure}
\end{figure}

An analysis of population exposure reveals striking disparities in the human impact of these wildfire events (Figure \ref{fig:pop_exposure}). The Palisades Fire had the greatest impact on the largest population, with 20,870 individuals exposed, while the Eaton Fire had the second-largest impact, affecting 20,193 people. Conversely, the Kenneth and Hurst Fires impacted substantially smaller populations, with 489 and 148 people exposed, respectively. This variation in population exposure can be attributed to the differential proximity of the fires to urban and rural areas.

\begin{figure}[h!]
    \centering
    \includegraphics[width=\textwidth]{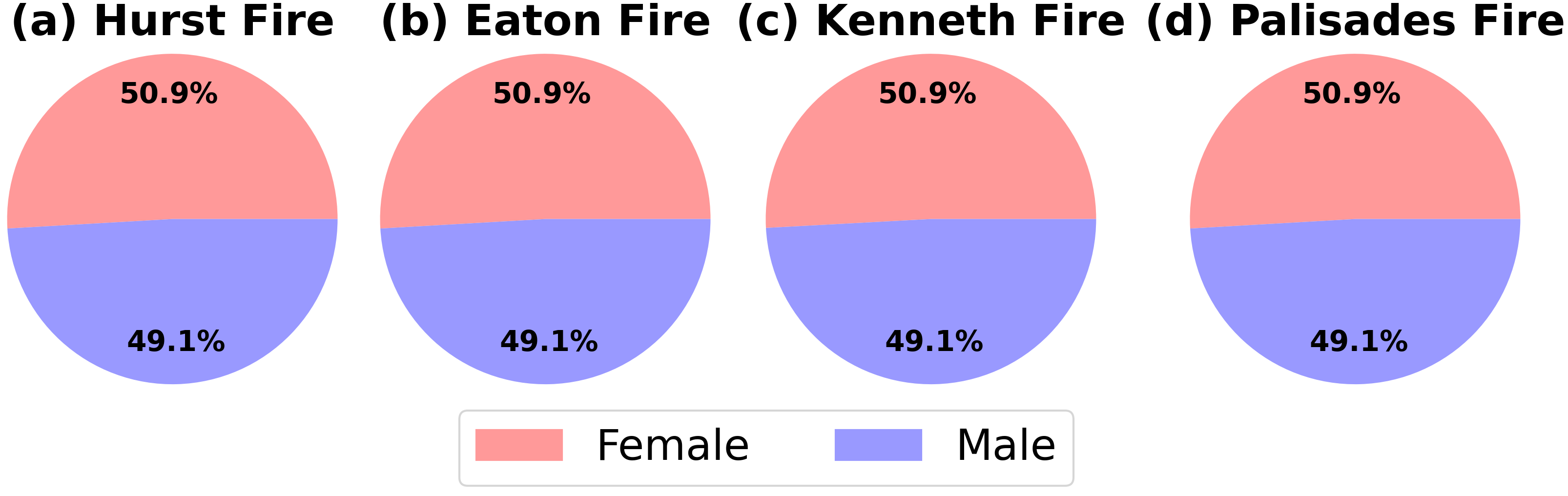}
    \caption{Gender distribution analysis across the four wildfire events, showing remarkably consistent patterns with approximately 50.9\% female and 49.1\% male population exposure across all fire incidents.}
    \label{fig:gender_dist}
\end{figure}

A thorough examination of the gender distribution across all four fire events reveals a strikingly consistent pattern (Figure \ref{fig:gender_dist}). Each fire zone exhibited nearly identical gender ratios, with females comprising 50.9\% and males 49.1\% of the affected population. This uniform distribution suggests that these fires affected communities with similar demographic compositions, despite their varying geographical locations and total population sizes.

\begin{figure}[h!]
    \centering
    \includegraphics[width=\textwidth]{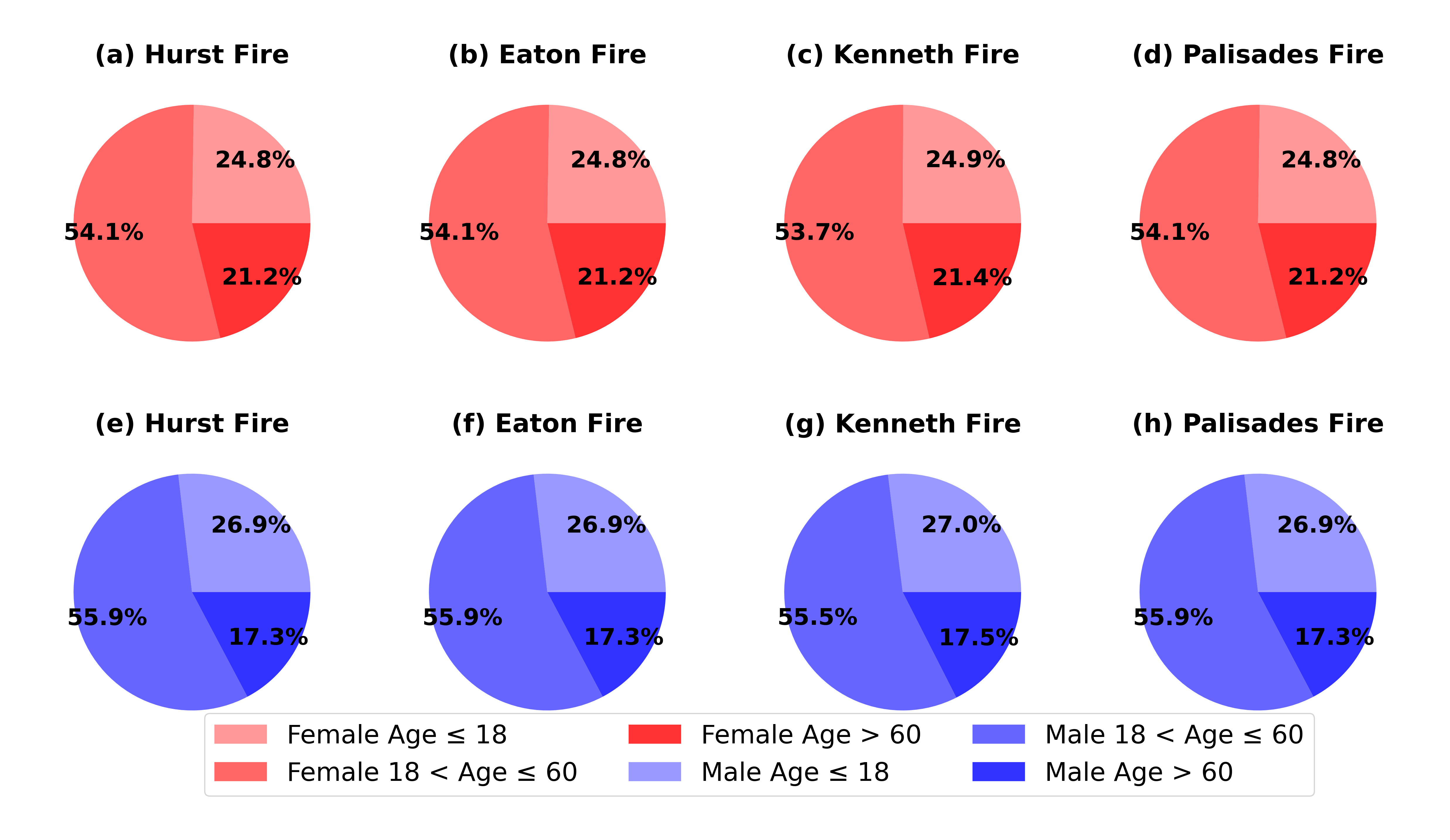}
    \caption{Combined age-gender distribution analysis across the four wildfire events: (a-d) female distribution across age groups in red tones, and (e-h) male distribution in blue tones.}
    \label{fig:age_gender_dist}
\end{figure}

A detailed age-gender distribution analysis reveals distinct demographic patterns between male and female populations across these fire events. The female population distribution (illustrated in Figures~\ref{fig:age_gender_dist}a--d, presented in red tones) shows consistent patterns across all fire events, with working-age women (18-60 years) constituting the predominant segment at approximately 54\% of the female population. The proportion of young females ($\leq$18 years) ranges from 24.8\% to 24.9\%, while elderly females ($>$60 years) constitute approximately 21.2\% to 21.4\% of the female population.

The male population (Figures~\ref{fig:age_gender_dist}e--h, shown in blue tones) exhibits a slightly different distribution pattern. Working-age men (18-60 years) demonstrate a marginally higher representation at 55.5-55.9\% compared to their female counterparts. Male youth ($\leq$18 years) constitute 26.9-27.0\% of the male population, while elderly males ($>$60 years) represent a smaller proportion at 17.3-17.5\%. This gender comparison reveals notable differences, particularly in the youth and elderly categories, with males having a higher proportion of youth (approximately 27\% vs 25\%) but a lower proportion of elderly individuals (approximately 17\% vs 21\%) compared to females.

The uniformity of these distributions across all four fire events, despite their diverse geographical locations, indicates consistent gender-age demographic patterns in these California communities. These findings carry significant ramifications for emergency response planning that must be sensitive to gender and age-specific needs. Particular attention should be paid to the higher proportion of elderly females and young males in these communities, while recognizing that working-age adults of both genders constitute the majority of the population. Emergency response strategies should account for these demographic variations to ensure appropriate resource allocation and support services for all age-gender groups.

\section{Discussion and Conclusions}

A thorough investigation of four significant California wildfires employing the Chebyshev-Kolmogorov-Arnold network (Cheby-KAN) model and multi-dimensional spatial analysis unveils intricate interactions between land cover, jurisdictional management, structural damage, and demographic vulnerability. The Cheby-KAN model exhibited a commendable degree of efficacy, with an overall accuracy of 98.68\%, a kappa coefficient of 0.9714, and an F1 score of 0.9817. This model's reliability is further substantiated by its capacity to generate precise burned area maps across a wide array of geographical landscapes. This high level of accuracy enabled precise quantification of fire extents ranging from 315.36 to 10,960.98 hectares, with the model effectively distinguishing between burned and unburned areas even in complex terrain.

The analysis of land cover, employing high-resolution (30m) National Land Cover Database (NLCD) data, revealed a conspicuous vulnerability of shrubland ecosystems. Tropical and Temperate Shrubland accounted for 57.4–75.8\% of burned areas across all fire events. This consistent pattern suggests that shrubland ecosystems serve as primary drivers of fire propagation due to their fuel characteristics and drought susceptibility. Spectral analysis of Sentinel-2 imagery (bands B2-B8, B11-B12) revealed distinct burn severity patterns, with the most severe burns occurring in areas of dense shrubland vegetation. The variation in secondary burn patterns, ranging from Mixed Forest (26.6\% in Hurst Fire) to Urban interfaces (19.0\% in Eaton Fire), demonstrates the complex fire behavior at ecosystem boundaries.

A jurisdictional analysis was conducted using the Protected Areas Database (PAD-US v2.0), which revealed significant variations in management complexity. The Palisades Fire exhibited a high degree of concentrated management (98.7\% single jurisdiction), while the Hurst Fire demonstrated distributed authority across the US Forest Service (31.9\%), Regional (15.6\%), and other agencies. This jurisdictional diversity had a direct impact on response coordination and resource allocation effectiveness. The analysis of demographic data from WorldPop, conducted at a resolution of 100 meters, revealed remarkably consistent gender distributions (50.9\% female, 49.1\% male) and age patterns across all events, with working-age populations (53.7-54.1\%) predominating.

Building infrastructure assessment using high-resolution California Building Footprints data uncovered stark urban-rural impact disparities. Urban interface fires (Eaton: 9,869 structures; Palisades: 8,436 structures) resulted in a significantly higher incidence of structural damage compared to rural events (Kenneth: 24 structures; Hurst: 17). This phenomenon was found to be strongly associated with population exposure metrics, with urban fires impacting over 20,000 individuals compared to less than 500 in rural areas. Spatial analysis revealed that structural damage clustered along wildland-urban interfaces, particularly in areas with high building density and complex topography.

These findings offer significant implications for technological advancement and policy development in wildfire management. The successful application of the Cheby-KAN model demonstrates the potential for machine learning approaches in rapid and accurate fire mapping. Furthermore, the observed patterns in ecological vulnerability and structural damage distribution underscore the necessity for location-specific management strategies that take into account both environmental and built environment characteristics.

Future research directions should focus on two key areas:(1) Development of predictive models that integrate demographic and structural vulnerability metrics for improved risk assessment, and (2) Investigation of temporal changes in burn patterns and severity under varying climate scenarios. Furthermore, future studies should explore the integration of LiDAR data for improved vegetation structure characterization and fire behavior modeling in complex terrain.

\bibliographystyle{unsrt}

\end{document}